\documentclass[twocolumn,floatfix,superscriptaddress,aps,prb]{revtex4}
\usepackage{epsf} 
\usepackage{times}
\usepackage{bm}
\usepackage{color}

\advance\textheight by 0.1in
\begin{document}

\title{Effect of anharmonicity of the strain energy on band offsets in
  semiconductor  nanostructures}

\author{Olga L. Lazarenkova} 
\email{Olga.L.Lazarenkova@jpl.nasa.gov}
\affiliation{Jet Propulsion Laboratory, California Institute of
  Technology, Pasadena CA 91109}

\author{Paul von Allmen}
\affiliation{Jet Propulsion Laboratory, California Institute of
  Technology, Pasadena CA 91109}

\author{Fabiano Oyafuso}
\affiliation{Jet Propulsion Laboratory, California Institute of
  Technology, Pasadena CA 91109}

\author{Seungwon Lee}
\affiliation{Jet Propulsion Laboratory, California Institute of
  Technology, Pasadena CA 91109}

\author{Gerhard Klimeck}
\affiliation{Jet Propulsion Laboratory, California Institute of
  Technology, Pasadena CA 91109}
\affiliation{Network for Computational Nanotechnology, School of
  Electrical and Computer Engineering, Purdue University, West
  Lafayette, IN 47906}

 \date{May 25, 2004}

 \begin{abstract} 
   Anharmonicity of the inter-atomic potential is taken into account
   for the quantitative simulation of the conduction and valence band
   offsets for highly-strained semiconductor heterostructures.  The
   anharmonicity leads to a weaker compressive hydrostatic strain than
   that obtained with the commonly used quasi-harmonic approximation
   of the Keating model.  Inclusion of the anharmonicity in the
   simulation of strained InAs/GaAs nanostructures results in an
   improvement of the electron band offset computed on an atomistic
   level by up to 100~meV compared to experiment.
 \end{abstract} 
 
 \maketitle
 
 The accurate simulation of the electronic structure is of utmost
 importance for the design of nanoelectronic and optoelectronic device
 structures. It has been shown both theoretically
 \cite{Shumway-Zunger, Pryor} and experimentally \cite{Shumway-Zunger,
   A3, A8, A7}, that the energy spectrum in semiconductor
 nanostructures is extremely sensitive to the built-in strain. Pryor
 {\it et al.}\ \cite{Pryor} has shown that the continuum elasticity
 method fails to adequately describe the strain profile in InAs/GaAs
 heterostructures with a 7\% lattice mismatch between the constituent
 materials. The two-parameter valence-force-field (VFF) Keating model
 \cite{Keating_VFFM, Martin_ZnS_VFFM} is a commonly used approximation
 for atomistic-level calculations of the equilibrium atomic positions
 in realistic-size nanostructures \cite{Zunger-Keating}. This paper
 shows that the quasi-harmonic Keating model is insufficient to
 describe highly strained InAs/GaAs nanostructures, where
 anharmonicity of the strain energy becomes important.

 Keating's model treats atoms as points connected with springs in a
 crystal lattice. The strain energy depends only the nearest-neighbor
 interactions within a quasi-harmonic approximation
 \cite{Keating_VFFM, Martin_ZnS_VFFM}:
\begin{eqnarray}
\label{E_s}
E&=&{3\over 8}\sum_m \biggl\{\sum_n \Bigl[ {\alpha_{mn}\over d_{mn}^2}({\bf
    r}_{mn} \cdot {\bf r}_{mn}-{\bf
    d}_{mn} \cdot {\bf d}_{mn})^2
\nonumber\\
&&\!\!\!\!\!\!\!\!\!+\sum_{k>n}{\sqrt{\beta_{mn}
    \beta_{mk}}\over d_{mn} d_{mk}}({\bf
    r}_{mn} \cdot {\bf r}_{mk}-{\bf
    d}_{mn} \cdot {\bf d}_{mk})^2\Bigr]\biggr\}.
\end{eqnarray}
The first coefficient, $\alpha$, corresponds to the spring constant
for the bond length distortion, while the second one, $\beta$,
corresponds to the change of the angle between the bonds or,
so-called, ``bond-bending''.  Here the summation is over all $m$ atoms
of the crystal and their nearest neighbors $n$ and $k$. ${\bf r}_{mn}$
and ${\bf d}_{mn}$ are the vectors connecting the $m$-th atom with its
$n$-th neighbor in the strained and unstrained material, respectively.

\begin{figure}[htb]
  \begin{center}
    \epsfbox{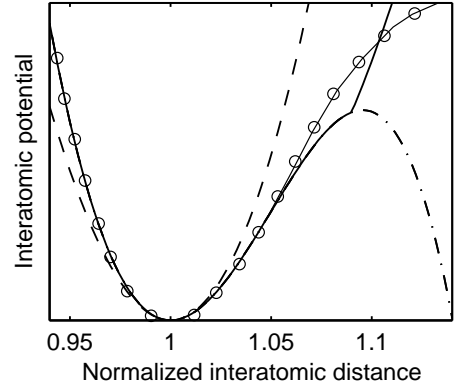}
    \caption{
      Schematic interatomic potential used in the Keating (dashed line)
      and our model (solid line).  Dash-dot line plot the model
      interatomic potential with the anharmonicity corrections to the
      VFF constants before the truncation. The line marked with large
      circles approximatly traces the shape of the realistic
      interatomic potential.}
    \label{Fig_E_dr_tt}
  \end{center}
\end{figure}
A realistic interatomic potential is schematically shown in
Fig.~\ref{Fig_E_dr_tt} with a line marked by large circles. 
The Keating interatomic potential corresponding to the Eq.~(\ref{E_s})
is plotted in Fig.~\ref{Fig_E_dr_tt} with dashed lines. The Keating
potential is referred to as ``quasi-harmonic'' rather  than
``harmonic'' since it depends upon the fourth rather  than the second
power of the bond length.  This quasi-harmonic potential fails to
reproduce the weakening of the interatomic interaction with 
increasing distance between atoms and it underestimates the repulsive
forces at close atomic separation. Therefore Eq.~(\ref{E_s}) can
adequately describe the strain energy only at small deformations. In
the widely used InAs/GaAs heterostructures, the lattice mismatch is
as large as 7\% and anharmonicity of the interatomic potential is
expected to become important.

We include the anharmonicity of the interatomic potential directly
into the Keating model by introducing distance-dependent VFF
constants $\alpha$ and $\beta$:
\begin{eqnarray}
\label{alpha}
\hskip-0.3in\alpha_{mn}&=&\alpha^{mn}_0\Bigl
            [1-A_{mn}{(r_{mn}^2-d_{mn}^2)\over d_{mn}^2}\Bigr],\\
\label{beta}
\hskip-0.3in\beta_{mnk}&=&
\beta_0^{nmk}
\Bigl[1-B_{nmk}{(\cos
    \theta^{nmk}-\cos \theta_0^{nmk})}\Bigr]
\nonumber\\
&&\times\Bigl[1-C_{nmk}{(r_{mn}r_{mk}-d_{mn}d_{mk})\over d_{mn}d_{mk}}\Bigr],
\end{eqnarray}
with the coefficients
$\beta_0^{nmk}\equiv\sqrt{\beta^{mn}_0\beta^{mk}_0}$, $B_{nmk}\equiv
\sqrt{B_{mn}B_{mk}}$, and $C_{nmk}\equiv \sqrt{C_{mn}C_{mk}}$.
$\theta^{nmk}$ and $\theta_0^{nmk}$ are the actual and the unstrained
angles between $mn$ and $mk$ bonds, respectively. In homogeneous
mono-atomic and binary compounds, all bonds are the same and the
indexes $m$, $n$, and $k$ can be dropped. $\alpha_0$ and $\beta_0$ are
the VFF constants in the unstrained material. The anharmonicity
correction coefficients $A$ and $C$ describe the dependence of
bond-stretching and bond-bending forces on hydrostatic strain, while
$B$ is responsible for the change of the bond-bending term with the
angle between bonds.  The details of the derivation of the
anharmonicity corrections $A$, $B$, and $C$ from the experimental
phonon spectra of strained bulk materials are presented in
Ref.~\cite{ase-prb}.  The parameters used for the simulation of
InAs/GaAs nanostructures are listed in Table~\ref{Tab_param}.
\begin{table}[htb]
  \begin{center}
    \begin{tabular}[c]{c|c c|c c c}
      \hline
      \hline
      $\quad$Material$\quad$ &$\quad$ $\alpha_0$ (N/m)&$\quad$ $\beta_0$ (N/m)$\quad$&$\;$ $A$$\;$&$\;$$B$$\;$ & $\;$$C$$\;$\\
      \hline
       GaAs & 41.49 & 8.94 & 7.2 & 7.62 & 6.4 \\
       InAs & 35.18 & 5.49 & 7.61 & 4.78 & 6.45   \\
      \hline
    \end{tabular}
    \caption{
      Valence-force-field constants in unstrained materials and
      anharmonicity correction coefficients for InAs and GaAs.}
    \label{Tab_param}
  \end{center}
\end{table}

Introduction of the anharmonicity corrections in the VFF expression
makes the form of the model potential more realistic and expands the
validity range of the strain simulations (see Fig.~\ref{Fig_E_dr_tt}).
In order to ensure the convergence of the minimization of the energy
(\ref{E_s}) with $\alpha$ and $\beta$ given by Eqs.~(\ref{alpha}),
(\ref{beta}), our model inter-atomic potential was truncated as shown
in Fig.~\ref{Fig_E_dr_tt} with the solid line.  Note that the proposed
introduction of the anharmonicity corrections in VFF constants does
not substantially increase the computational time of the minimization
of the total strain energy, since it does not introduce any additional
summation in Eq.~(\ref{E_s}).

\begin{table*}[htb]
  \begin{center} \begin{tabular}[c]{c | c c | c c c c| c c c | c c c }
  \hline\hline Ref. & $x$ & $y$ & Structure &
  \multicolumn{2}{c}{Size}& Substr. & \multicolumn{3}{c|}{Hydrostatic
  strain (\%)} & \multicolumn{3}{c}{Biaxial strain (\%)}\\
  &&&&$\quad$$L_x$$\quad$&$\quad$$L_y$$\quad$&&
  $\quad$$\epsilon_H^K$$\quad$& $\quad$$\epsilon_H^A$$\quad$ &
  $\quad$$\delta_H$$\quad$& $\quad$$\epsilon_B^K$$\quad$&
  $\quad$$\epsilon_B^A$$\quad$ & $\quad$$\delta_B$$\quad$ \\ 
        \hline
  \cite{A3}& 1.0 & 0.0& SQW &2~ML& 5~ML 
  &GaAs & -2.97 & -2.59 & 14.9 & -3.71 & -4.10 & -9.4 \\ 
  \cite{A3}& 0.0 & 1.0& SQB &2~ML& 5~ML
  &InAs & 3.54 & 4.30 & -17.7 & 3.72 & 2.90 & 28.3\\ 
  \cite{A8}& 1.0 & 0.0& MQW &1~ML&30~nm 
  &GaAs & -2.87 & -2.33 & 23.5 & -3.81 & -4.38 & -13.0 \\ 
        \hline 
 \end{tabular} 
  \caption{ 
    Hydrostatic (H) and biaxial (B) strain components within the thin
    epitaxial layer in several structures computed within the Keating
    (K) model and taking into account the anharmonicity of the strain
    energy (A) for different In$_x$Ga$_{1-x}$As/In$_y$Ga$_{1-y}$As
    nanostructures \cite{A3, A8}.  {\it Notations}: the nanostructure
    type: SQW -- single quantum well, SQB -- single quantum barrier,
    MQW -- multiple quantum wells, ML -- monolayer, $L_x$ -- width of
    the In$_x$Ga$_{1-x}$As epitaxial layer, $L_y$ -- width of the
    In$_y$Ga$_{1-y}$As epitaxial layer (corresponds to the thickness
    of the capping layer for SQW and SQB).
    $\delta=(\epsilon^K-\epsilon^A)/\epsilon^A$ shows the relative
    change of the computed built-in strain when the anharmonicity is
    neglected.  }
\label{Tab_eps} 
\end{center}
\end{table*}
To illustrate the effect of the anharmonicity on the strain
distribution in III-V semiconductor nanostructures, the hydrostatic,
$\epsilon_H=1/3(\epsilon_{xx}+\epsilon_{yy}+\epsilon_{zz})$, and
biaxial, $\epsilon_B=1/6(\epsilon_{xx}+\epsilon_{yy}-2\epsilon_{zz})$,
components of the strain in ultra-thin epitaxial InAs/GaAs layers have
been computed using both the conventional Keating model and our model.
The geometry of the structures and the obtained built-in strain are
presented in Table~\ref{Tab_eps}.
Comparing the results of the two models, we note that the sharp rise
of the strain energy at small inter-atomic distances leads to a
smaller equilibrium hydrostatic compression than is obtained with the
Keating model. At the same time, the quasi-harmonic approximation
underestimates the bond stretching deformation. In contrast, biaxial
compression is increased in our anharmonic model, while the biaxial
tension is suppressed.

\begin{table*}[htb]
  \begin{center}
    \begin{tabular}[c]{l | c |  c | c  c c c c | c c c  c  c  c  }
      \hline\hline
      Ref. & Structure & Experimental &
      \multicolumn{5}{c|}{$\Delta E_c$ (meV)} &
      \multicolumn{6}{c}{$\Delta E_v$ (meV)} \\
      &&method&
      $\quad$K$\quad$&$\delta_K$ (\%)$\quad$&
      $\quad$A$\quad$&$\delta_A$ (\%)$\quad$&$\quad$Exp.$\quad$&
      $\quad$K$\quad$&$\delta_K$ (\%)$\quad$&
      $\quad$A$\quad$&$\delta_A$ (\%)$\quad$&$\quad$Exp.$\quad$&
      Sub-band\\
      \hline
      \cite{A3}&SQW &XPS&471.5&&574.0&& &374.6 &-34.4&447.0 &-15.7&$530\pm50$&$hh$\\
      \cite{A3}&SQB &XPS&-40.7&&-91.3 && &-271.7& 69.8&-225.9&41.2&$-160\pm50$&$lh$\\
      \cite{A8}&MQW &CV\&DLTS&475.7&-31.1&584.4&-15.3& $690$&373.4&&430.5&&&$hh$ \\
      \cite{A7}&QDC & DLTS&242.0&-29.0&347.0&1.8&$341\pm 30$ &&&&&&  \\
      \hline
   \end{tabular}
    \caption{
      Experimental band offsets in the conduction ($\Delta E_c$) and
      valence ($\Delta E_v$) bands compared with the offsets computed
      within the $sp^3d^5s^*$ empirical tight-binding model
      \cite{NEMO3D} using the equilibrium atomic positions found
      within the two-parameter Keating model (K) and including
      anharmonicity corrections to the VFF constants (A) for different
      In$_x$Ga$_{1-x}$As/In$_y$Ga$_{1-y}$As nanostructures. The SQW,
      SQB, and MQW structures are the same as in Table~\ref{Tab_eps}.
      The QDC structure \cite{A7} consists of three layers of
      dome-shaped QDs with a 20~nm base and a 7~nm height on top of a
      0.7~nm wetting layer.  The distance between the QD layers is
      about 10~nm.  The band offsets are determined as the following:
      $\Delta E_c=E_c($In$_y$Ga$_{1-y}$As$)-E_c($In$_x$Ga$_{1-x}$As$)$
      and $\Delta
      E_v=-(E_v($In$_y$Ga$_{1-y}$As$)-E_v($In$_x$Ga$_{1-x}$As$))$ so
      they would be positive for potential well and negative for
      potential barrier.  {\it Notations}: the nanostructure type: SQW
      -- single quantum well, SQB -- single quantum barrier, MQW --
      multiple quantum wells, QDC -- quantum dot crystal, i.e., the
      3-dimensional {\it ordered} array of quantum dots; the band
      offset measurement techniques: XPS -- X-ray photoemission
      spectroscopy, CV -- capacitance-voltage spectroscopy, DLTS --
      deep-level transient spectroscopy. $\delta_K$ and $\delta_A$
      estimate the relative deviations of the simulation from the
      experiment.  }
    \label{Tab_InGaAs}
  \end{center}
\end{table*}
The band offsets for different InAs/GaAs nanostructures obtained for
the strain distribution simulated within the Keating's model and
including the anharmonicity corrections are compared with the
available experimental data \cite{A3, A8, A7} in
Table~\ref{Tab_InGaAs}. The local band structure was obtained within
the $sp^3d^5s^*$ empirical tight-binding model where the Hamiltonian
matrix elements depend on the distance between the atoms
\cite{NEMO3D}. The tight-binding parameters were fitted to reproduce
the properties both of the unstrained and strained bulk materials.
Ultra-thin epitaxial layer structures \cite{A3, A8} provide us with
the perfect opportunity to test the effect of the anharmonicity at
strong deformation on the valence and conduction band offsets.
Anharmonicity corrections reduce the discrepancy between the
experimental and modeled energies significantly (see
Table~\ref{Tab_InGaAs}).

\begin{figure}[htb]
  \begin{center}
    \epsfbox{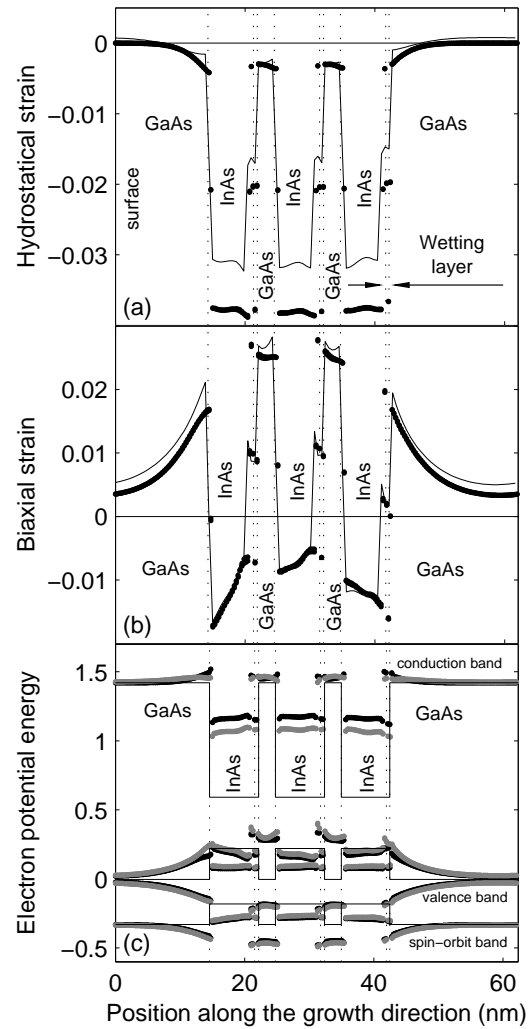}
    \caption{  
      Computed distribution of the hydrostatic (a) and biaxial (b)
      strain components, and (c) electronic band structure along the
      growth direction in InAs/GaAs QDC.  The structure taken from
      Ref.~\cite{A7} consists of three layers of dome-shaped QDs with
      a 20~nm base and a 7~nm height on top of the 0.7~nm wetting
      layer.  The crossection is made near the center of the quantum
      dot stack.  The results obtained with the Keating model are
      plotted with black dots. The results obtained with the
      anharmonic model are plotted with solid line on (a) and (b) and
      with gray dots on (c). The thin lines on (c) show the edges of
      conduction, valence, and spin-orbit split-off bands at the
      center of the Brillouin zone in the unstrained materials.}
    \label{Fig_qdc}
  \end{center}
\end{figure}
The strain distribution (Fig.~\ref{Fig_qdc}, a, b) and the energy
spectrum (Fig.~\ref{Fig_qdc}, c) were computed for the quantum dot
crystal (QDC) reported in Ref.~\cite{A7}. The structure consists of
three layers of {\it regimented} dome-shaped quantum dot arrays (with
a 20~nm base diameter and a 7~nm height), vertically stacked with a
small (3~nm) vertical separation between the QD layers. The 0.7~nm
wetting layer is also included in our simulations to properly model
the electronic spectrum \cite{WL}.  The built-in strain distribution
in such structures is very inhomogeneous.  The average hydrostatical
component of the strain tensor inside InAs quantum dots
(Fig.~\ref{Fig_qdc}, a) is overestimated by about 25\% within the
commonly used Keating model.  At the same time, the biaxial strain
distribution (Fig.~\ref{Fig_qdc}, b) changes little when computed with
the different models.

The main effect of the anharmonicity is a downward shift of the
conduction band edge inside the quantum dots, as shown in
Fig.~\ref{Fig_qdc}, c.  This is caused by the sensitivity of the
conduction band to the hydrostatic compression, which is smaller
within the anharmonicity model. The difference in the corresponding
band edges obtained within the two models is as large as 105.0~meV.
This shift brings the overall band offset between the InAs quantum dot
and GaAs buffer computed in our model very close to the experimentally
observed value (see Table~\ref{Tab_InGaAs}). Due to the small
difference in the biaxial strain distribution obtained with the two
models (see Fig.~\ref{Fig_qdc}, b), the energy structure of the
valence band remains almost the same, as can be seen in
Fig.~\ref{Fig_qdc}, c.

In conclusion, we have demonstrated that the anharmonicity is
important for the modeling of the electronic states in highly strained
InAs/GaAs system. Compared to the standard Keating model we have found
corrections of over 100~meV in some band offsets, resulting in values
significantly closer to the experimental data. This demonstrates that
the deformation in the nanostructures is beyond the applicability
range of the quasi-harmonic approximation for the strain energy. The
anharmonicity corrections can be performed without a significant
increase of the computational cost, since the model remains limited to
a nearest neighbour interactions.

{\it Aknowledgements. } The work described in this publication was
carried out at the Jet Propulsion Laboratory, California Institute of
Technology under a contract with the National Aeronautics and Space
Administration.  Funding was provided under grants from ARDA, ONR,
JPL, NASA and NSF (Grant No.\ EEC-0228390).  This work was performed
while one of the authors (O.L.L.) held a National Research Council
Research Associateship Award at JPL.

\end{document}